\documentclass[pdflatex,sn-mathphys-num]{sn-jnl}

\usepackage{graphicx}%
\usepackage{multirow}%
\usepackage{amsmath,amssymb,amsfonts}%
\usepackage{amsthm}%
\usepackage{mathrsfs}%
\usepackage[title]{appendix}%
\usepackage{xcolor}%
\usepackage{textcomp}%
\usepackage{manyfoot}%
\usepackage{booktabs}%
\usepackage{algorithm}%
\usepackage{algorithmicx}%
\usepackage{algpseudocode}%
\usepackage{listings}%
\usepackage{subfigure}
\usepackage{tikz}
\usepackage{amsmath}
\usepackage[all, ps, dvips]{xy} 
\usepackage{qcircuit} 
\usepackage{braket}

\theoremstyle{thmstyleone}%
\theoremstyle{thmstyletwo}%

\theoremstyle{thmstylethree}%

\begin{document}

\title[Article Title]{Exploring the Potential of QEEGNet for Cross-Task and Cross-Dataset Electroencephalography Encoding with Quantum Machine Learning}


\author*[1]{\fnm{Chi-Sheng} \sur{Chen}}\email{m50816m50816@gmail.com}

\author[2]{\fnm{Samuel Yen-Chi} \sur{Chen}}\email{ycchen1989@ieee.org}

\author[2]{\fnm{Huan-Hsin} \sur{Tseng}}\email{
htseng@bnl.gov}

\affil*[1]{\orgname{Neuro Industry, Inc.}, \orgaddress{\city{Boston}, \state{MA}, \country{USA}}}

\affil[2]{\orgdiv{AI Department}, \orgname{Brookhaven National Laboratory}, \orgaddress{\city{Upton}, \state{NY}, \country{USA}}}



\abstract{Electroencephalography (EEG) is widely used in neuroscience and clinical research for analyzing brain activity. While deep learning models such as EEGNet have shown success in decoding EEG signals, they often struggle with data complexity, inter-subject variability, and noise robustness. Recent advancements in quantum machine learning (QML) offer new opportunities to enhance EEG analysis by leveraging quantum computing’s unique properties. In this study, we extend the previously proposed Quantum-EEGNet (QEEGNet), a hybrid neural network incorporating quantum layers into EEGNet, to investigate its generalization ability across multiple EEG datasets. Our evaluation spans a diverse set of cognitive and motor task datasets, assessing QEEGNet’s performance in different learning scenarios. Experimental results reveal that while QEEGNet demonstrates competitive performance and maintains robustness in certain datasets, its improvements over traditional deep learning methods remain inconsistent. These findings suggest that hybrid quantum-classical architectures require further optimization to fully leverage quantum advantages in EEG processing. Despite these limitations, our study provides new insights into the applicability of QML in EEG research and highlights challenges that must be addressed for future advancements.
}




\maketitle

\section{Introduction}\label{sec1}

Electroencephalography (EEG) plays a fundamental role in neuroscience and clinical applications, enabling non-invasive monitoring and analysis of brain activity \cite{chen2025lcm}. While deep learning models such as EEGNet \cite{lawhern2018eegnet} have demonstrated substantial progress in decoding EEG signals, they often encounter limitations when faced with the high dimensionality, variability, and noise inherent in EEG data. These challenges motivate the exploration of novel computational paradigms that can enhance EEG representation and classification. Emerging research has demonstrated the potential of EEG in diverse applications beyond traditional medical diagnostics. For instance, EEG signals have been successfully integrated into real-time robotic control systems, as seen in \cite{chen2024psycho}. Furthermore, EEG has been utilized in multimodal frameworks, such as \cite{chen2024necomimi}, where EEG signals serve as a guiding input for generating images, pushing the boundaries of EEG’s creative applications. Similarly, \cite{chen2024mind} has explored EEG-based visual recognition tasks by leveraging contrastive learning to align EEG and visual representations. In clinical neuroscience, EEG has also been explored as a predictive tool for psychiatric treatment responses \cite{li2023prediction}. These studies highlight EEG’s versatility in various AI-driven applications and the increasing demand for advanced models capable of capturing intricate brain activity patterns.

Quantum machine learning (QML) has emerged as a promising field that leverages quantum computing's unique properties \cite{biamonte2017quantum}, such as superposition and entanglement, to potentially improve learning efficiency and model expressivity \cite{abbas2021power,du2020expressive,caro2022generalization}. 
Recent advancements in variational quantum circuits (VQCs) have demonstrated their capability in deep reinforcement learning \cite{chen2020variational}, long short-term memory (LSTM) architectures \cite{chen2022quantum}, and transfer learning \cite{tseng2025transfer}. These developments suggest that QML-based models can potentially enhance learning in complex, high-dimensional spaces by leveraging quantum-native representations. Moreover, studies have investigated the role of quantum measurement in optimizing quantum neural networks \cite{chen2025learning}, shedding light on techniques to improve the trainability and interpretability of hybrid quantum-classical models. Another notable work introduces Quantum-Train Agents for programming variational quantum circuits, further advancing QML methodologies \cite{liu2024programming}.
Notably, \cite{chen2024quantum} has investigated the potential of quantum-enhanced contrastive learning in fusing EEG and image representations, illustrating that QML can be leveraged beyond unimodal EEG analysis. In our prior work \cite{chen2024qeegnet}, Quantum-EEGNet (QEEGNet) was introduced as a hybrid neural network that integrates quantum layers within the classical EEGNet framework. Initial findings indicated that QEEGNet could capture intricate patterns in EEG signals and demonstrated competitive performance on benchmark EEG datasets, such as BCI Competition IV 2a \cite{tangermann2012review}. However, its effectiveness across broader datasets and real-world scenarios remains an open question. In this study, we extend our previous work by systematically evaluating QEEGNet on multiple public EEG datasets encompassing a range of cognitive and motor tasks. This investigation aims to assess its generalization ability and robustness beyond a single benchmark dataset. Specifically, we explore whether QEEGNet's quantum-enhanced representation can provide advantages in EEG decoding tasks involving different levels of data complexity, noise, and inter-subject variability. Our results indicate that while QEEGNet exhibits promising trends in certain datasets, its performance does not always surpass traditional deep learning approaches. This underscores the need for further optimization of quantum-classical hybrid architectures and a deeper understanding of how quantum layers interact with EEG feature extraction. Nonetheless, our findings contribute to the growing body of research on QML applications in neuroscience, highlighting both the opportunities and challenges of integrating quantum computing into EEG analysis.

\begin{figure}
    \centering
    \includegraphics[width=1\linewidth]{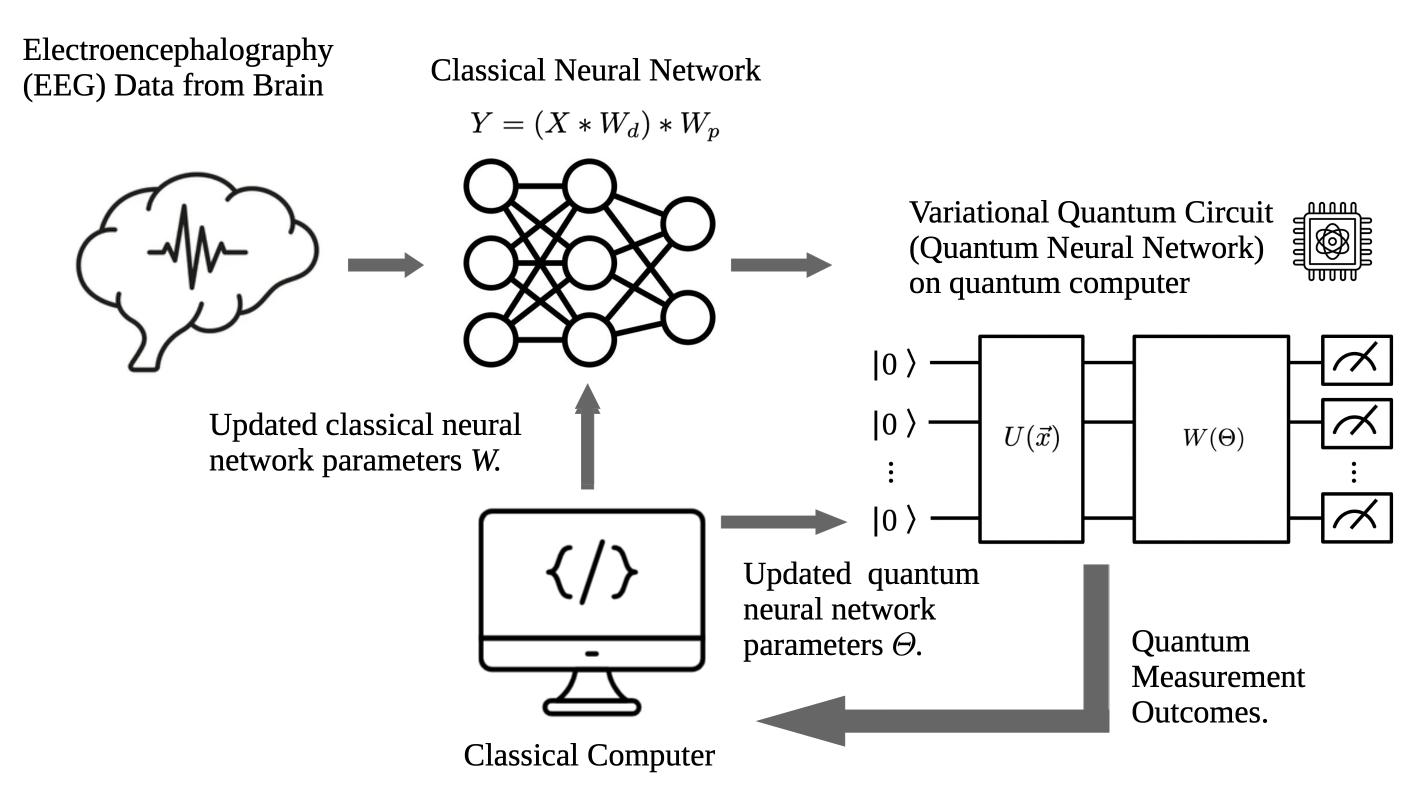}
    \caption{Hybrid Quantum-Classical Framework for EEG Processing in QEEGNet. The framework integrates a classical neural network with a variational quantum circuit to process electroencephalography (EEG) data. EEG features are first extracted and transformed by a classical neural network, parameterized by $W$. The resulting feature representations are then encoded into a quantum state and processed using a quantum neural network with trainable parameters $\Theta$. The quantum measurement outcomes are fed back into a classical computer, updating both quantum and classical parameters iteratively. This hybrid approach leverages quantum computing to enhance feature representation learning for EEG-based applications.}
    \label{fig:imgabs}
\end{figure}

By expanding the scope of evaluation, this work provides insights into the potential and limitations of quantum-enhanced EEG models. We discuss key factors influencing QEEGNet’s performance, propose directions for improving hybrid architectures, and outline future research pathways toward practical applications of quantum computing in EEG-based brain-computer interfaces and clinical diagnostics. The whole structure is shown in Figure~\ref{fig:imgabs}. Our contributions in this paper are as follows:

\begin{itemize}
    \item Building upon our previous work, we propose QEEGNet, a quantum machine learning model specifically designed for EEG analysis.
    \item We conduct extensive evaluations across multiple commonly used EEG datasets, performing both cross-task and cross-dataset validation to assess the model's generalizability.
    \item We provide a detailed computational complexity analysis comparing QEEGNet with the traditional EEGNet, highlighting the trade-offs between quantum-enhanced feature representation and computational efficiency.
\end{itemize}

\section{Methodology}\label{sec2}

In this section, we present the methodology behind Quantum-EEGNet (QEEGNet), a novel hybrid neural network that integrates quantum computing principles with classical deep learning architectures to enhance EEG data encoding and classification, is shown in Fig.~\ref{fig:qeegnet}. Our approach builds upon the well-established EEGNet framework, incorporating quantum layers to leverage the computational advantages of quantum mechanics, such as superposition and entanglement. The overall model structure is designed to capture the intricate spatial and temporal dependencies present in EEG signals more effectively than purely classical architectures.
\begin{figure}[h]
    \centering
    \begin{tikzpicture}

        \draw[thick] (-2, 0) node[left] {$|0\rangle$} -- (-1.5, 0);
        \draw[thick] (-2, -1) node[left] {$|0\rangle$} -- (-1.5, -1);
        \draw[thick] (-2, -3) node[left] {$|0\rangle$} -- (-1.5, -3);

        \node at (-1.7, -2) {\vdots};

        \draw[thick] (0.5, 0) -- (1.5, 0);
        \draw[thick] (0.5, -1) -- (1.5, -1);
        \draw[thick] (0.5, -3) -- (1.5, -3);

        \draw[thick] (3, 0) -- (3.5, 0);
        \draw[thick] (3, -1) -- (3.5, -1);
        \draw[thick] (3, -3) -- (3.5, -3);

        \draw[thick] (5, 0) -- (5.5, 0);
        \draw[thick] (5, -1) -- (5.5, -1);
        \draw[thick] (5, -3) -- (5.5, -3);

        \node at (6.5, -2) {\vdots};

        \draw[thick] (-1.5, 0.5) rectangle (0.5, -3.5);
        \node at (-0.5, -1.5) {$U(\vec{x})$};
        \node[below] at (-0.5, -4) {\textbf{Encoding Block}}; 

        \draw[dashed, thick] (1, 0.8) rectangle (5.8, -3.8);
        \draw[thick] (1.5, 0.5) rectangle (3, -3.5);
        \draw[thick] (3.5, 0.5) rectangle (5, -3.5);
        \node at (2.25, -1.5) {$V_1(\vec{\theta}_1)$};
        \node at (4.25, -1.5) {$V_n(\vec{\theta}_n)$};
        \node[below] at (3.5, -4) {\textbf{Variational Block}}; 

        \node at (3.25, -1.5) {\dots};

        \foreach \y in {0,-1,-3} {
            \draw[thick] (5.5, \y) -- (6, \y);
            \draw[thick] (6.8, \y) -- (7, \y);

            \draw[thick] (6, \y+0.25) rectangle (6.8, \y-0.25);

            \draw[thick] (6.5, \y-0.1) arc (0:180:0.15); 

            \draw[thick] (6.7, \y+0.15) -- (6.1, \y-0.15); 
        }

        \node[below] at (6.5, -4) {\textbf{Measurement}}; 

    \end{tikzpicture}
    \caption{Structure of the Variational Quantum Circuit (VQC). The input $\vec{x}$ represents EEG features encoded by a classical neural network before being processed by the quantum circuit. The circuit consists of an encoding layer $U(C(\vec{x}))$, variational layers $V_i(\vec{\theta}_i)$, and measurement operations to extract quantum outputs.}

    \label{fig:vqc_diagram}
\end{figure}
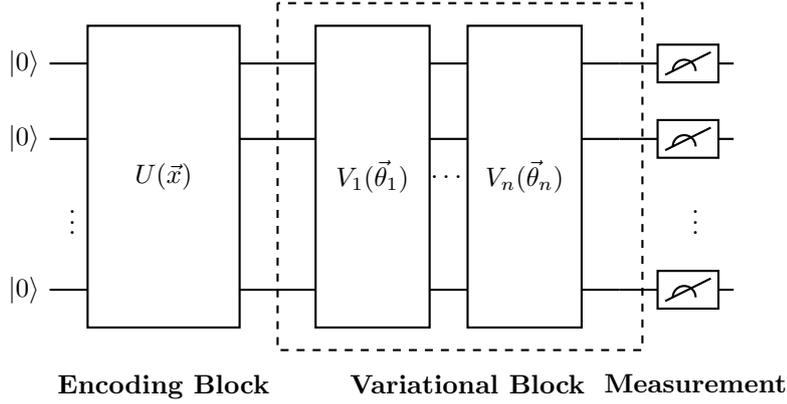

\subsection{Classical Feature Extraction}
EEGNet serves as the backbone of QEEGNet, initially processing raw EEG signals to extract relevant spatial and temporal features. The input EEG data, represented as a multichannel time-series matrix $X \in \mathbb{R}^{C \times T}$, where $C$ is the number of channels and $T$ is the number of time steps, undergoes a series of convolutional operations to model spatial and temporal dependencies. The first stage applies depthwise separable convolutions, given by:
\begin{equation}
Y = (X * W_d) * W_p,
\end{equation}
where $W_d$ represents the depthwise convolution kernel applied per channel, and $W_p$ denotes the pointwise convolution kernel used for cross-channel feature fusion.

Batch normalization follows to stabilize activations, ensuring improved gradient flow and faster convergence. The activation function utilized is the Exponential Linear Unit (ELU) \cite{clevert2015fast}, defined as:
\begin{equation}
ELU(x) =
\begin{cases}
x, & x > 0 \\
\alpha (e^x -1), & x \leq 0,
\end{cases}
\end{equation}

where $\alpha$ controls the saturation rate for negative inputs. By stacking multiple convolutional layers with ELU activations and pooling operations, the network efficiently compresses EEG representations while preserving discriminative features.

\subsection{Quantum Encoding Layer}

In the Quantum-Classical Hybrid Neural Network (HNN) architecture, a Classical Neural Network is combined with a Variational Quantum Circuit (VQC) which is shown in Figure~\ref{fig:vqc_diagram}, enabling classical data to undergo feature extraction before being transformed into quantum states, leveraging the advantages of quantum computation for further processing. The model is defined as:

\begin{equation}
    f_{\text{HNN}} : \mathbb{R}^d \to \mathbb{R}, \quad f_{\text{HNN}} = M \circ V \circ U \circ C
\end{equation}

 where $C: \mathbb{R}^d \to \mathbb{R}^{d'}$ is the classical neural network, mapping input data $x \in \mathbb{R}^d$ to a lower-dimensional embedding space 
with $d > d'$. A quantum embedding unitary $U: \mathbb{R}^{d'} \to \mathcal{U}(\mathcal{H})$ is defined to transform an embedded classical input $C(x)$ into a quantum state by $\ket{\psi} \mapsto U(C(x))\ket{\psi}$. Subsequently, we define a variational quantum circuit $V: \mathcal{U}(\mathcal{H}) \to \mathcal{U}(\mathcal{H})$ with tunable parameters $\theta$ to generate a corresponding quantum transformation $V(\theta)$. A final step of VQC is to perform a measurement on the resulting quantum state with a selected observable  $M: \mathcal{H} \to \mathbb{R}$. The computation process above combined yields,
\begin{equation}\label{E: output}
    f_{\text{HNN}}(x) = \langle \psi_0 | U^{\dagger}(C(x)) V^{\dagger}(\theta) H V(\theta) U(C(x)) | \psi_0 \rangle
\end{equation}
where $V^{\dagger}$ and $U^{\dagger}$ denote the Hermitian conjugates
of the encoding and variational quantum circuit, respectively. 

The classical embedding $C(x)$ is implemented by a classical neural network of multiple layers,
\begin{equation}
    C(x) = \sigma(W_L \sigma(W_{L-1} \dots \sigma(W_1 x + b_1) \dots + b_{L-1}) + b_L)
\end{equation}

where $W_i \in \mathbb{R}^{d_i \times d_{i-1}}$ and $b_i \in \mathbb{R}^{d_i}$ are the weight matrices and biases of the neural network, and $\sigma$ is a non-linear activation function (e.g., ReLU or Sigmoid). The output $C(x) \in \mathbb{R}^{d'}$ serves as an input to the quantum circuit. 

The variational quantum circuit consists of a sequence of parameterized unitary operations $V(\theta) = V_L(\theta_L) \cdots V_1(\theta_1)$ with $V_{\ell}(\theta_{\ell}) = e^{-\frac{i}{2} \theta_{\ell} \sigma_{\ell}}$ for $\ell = 1, \ldots, L$, and  $\sigma_{\ell}$ is one of the Pauli matrices,
\begin{equation}\label{E: Pauli}
    \sigma_x = \begin{pmatrix}
    0 & -1 \\
    1 & 0
    \end{pmatrix}, \quad  \sigma_y = \begin{pmatrix}
    0 & -i \\
    i & 0
    \end{pmatrix}, \quad \sigma_z = \begin{pmatrix}
    1 & 0 \\
    0 & -1
    \end{pmatrix}.
\end{equation}
depending on $\ell$. The measurement is performed by a selected Hermitian observable $H$ such that $M(|\psi\rangle) = \langle \psi | H | \psi \rangle \in \mathbb{R}$. A typical choice is the Pauli matrix in Eq.~(\ref{E: Pauli}).


The optimization over tunable parameters $\theta$ is
performed by minimizing the loss function:

\begin{equation}
    \mathcal{L}(\theta) = \frac{1}{N}  \sum_{i=1}^{N} \left( f_{\text{HNN}}(x^{(i)}; \theta) - y^{(i)} \right)^2
\end{equation}

This hybrid model integrates the feature learning capabilities of classical neural networks with the nonlinear mapping properties of quantum transformations, enabling superior feature extraction and modeling performance in specific problem domains.

\begin{figure}
    \centering
    \includegraphics[width=.8\linewidth]{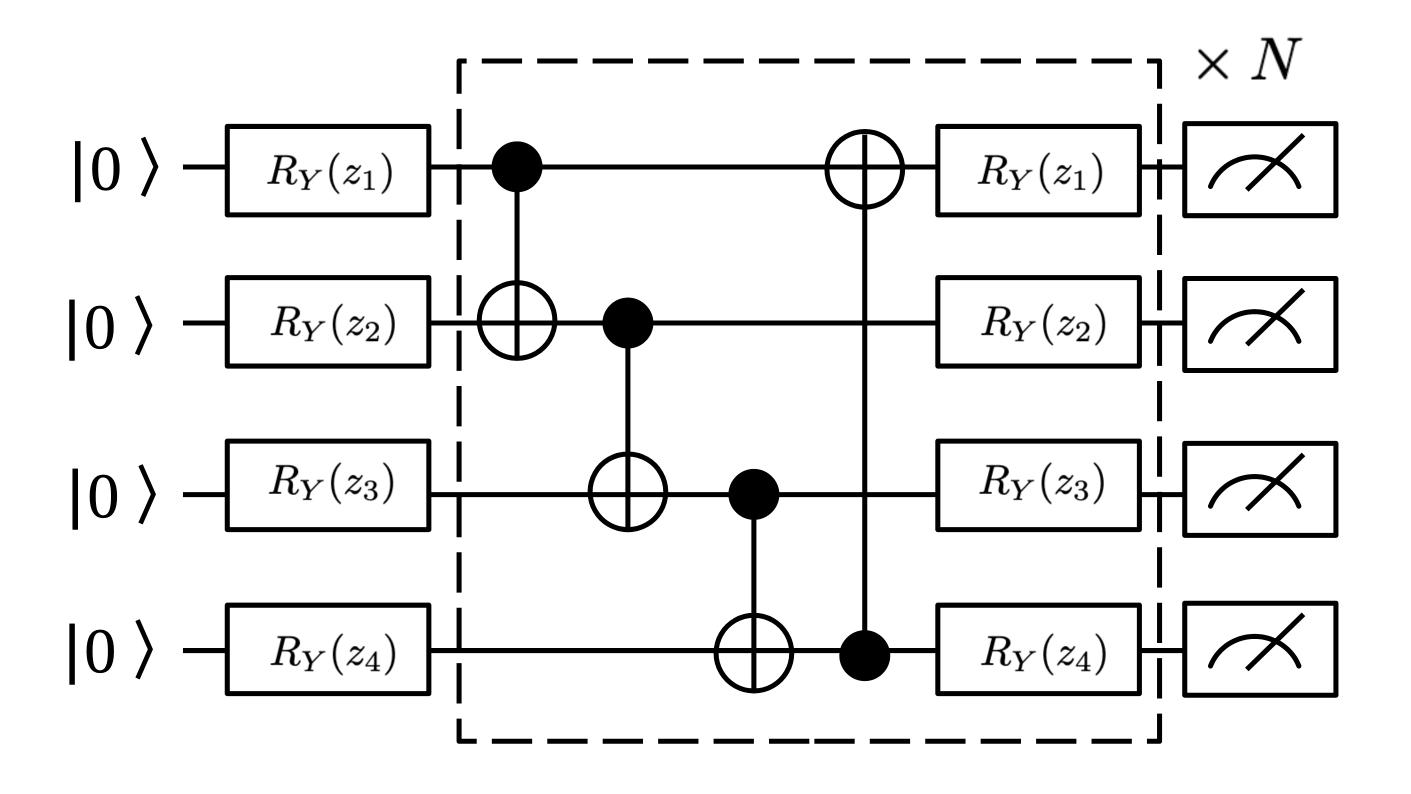}
    \caption{The VQC (QNN) layer in this work.}
    \label{fig:qeeg_qlayer}
\end{figure}

\begin{figure}
    \centering
    \includegraphics[width=1\linewidth]{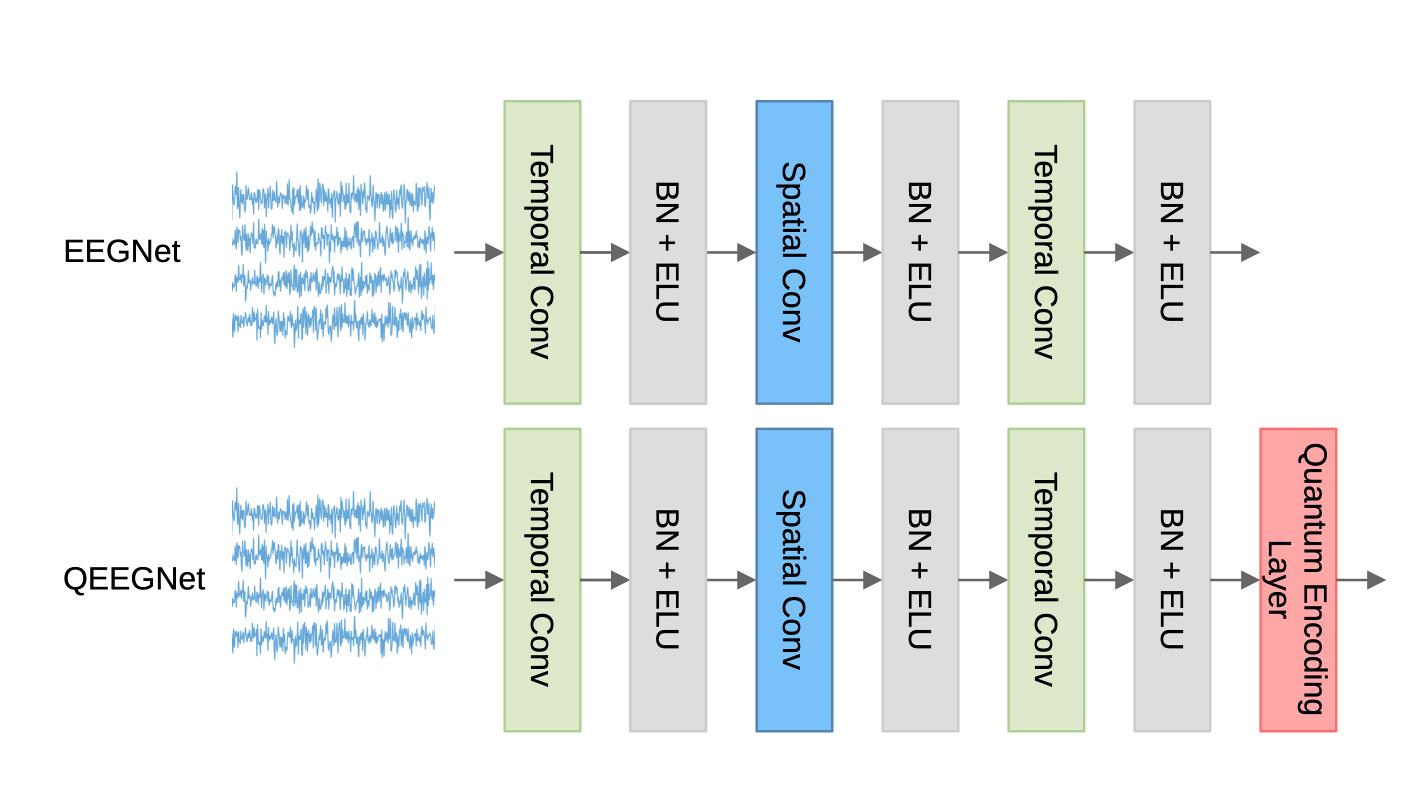}
    \caption{A schematic comparison between EEGNet and the proposed Quantum-EEGNet (QEEGNet). In this illustration, Conv represents convolution operations, BN stands for batch normalization, and ELU refers to the exponential linear unit activation function.}
    \label{fig:qeegnet}
\end{figure}

Once EEG features are extracted by the classical neural network, they are transformed into quantum representations through a series of quantum encoding operations. Each extracted feature vector $z = (z_1, \ldots, z_{d'}) \in \mathbb{R}^{d'}$ is mapped onto a quantum state via a parameterized rotation operation like Figure~\ref{fig:qeeg_qlayer}.

The encoding {$U$} utilizes a rotation-$Y$ (RY) gate:
\begin{equation}
R_Y(z) = \exp\left(-i z \frac{\sigma_y}{2}\right),
\end{equation}
where $\sigma_y$ is the Pauli-$Y$ matrix in Eq.~(\ref{E: Pauli}). With this transformation, classical data is embedded into quantum states of a Hilbert space for richer representations.

To introduce interdependencies among qubits, entanglement is established via a ring-structured application of controlled-NOT (CNOT) gates. The CNOT operation, acting on a control qubit $q_c$ and a target qubit $q_t$, is defined by the unitary matrix:
\begin{equation}
\text{CNOT} = \begin{pmatrix}
1 & 0 & 0 & 0 \\
0 & 1 & 0 & 0 \\
0 & 0 & 0 & 1 \\
0 & 0 & 1 & 0
\end{pmatrix}.
\end{equation}

Applying this gate across neighboring qubits ensures that quantum correlations between EEG features are preserved, effectively capturing complex dependencies that may be lost in purely classical representations. A ring entanglement pattern arranges CNOT gates in a way that each qubit is entangled with its adjacent neighbor, while the final qubit is also entangled with the first qubit, creating a \textbf{closed-loop structure}. This setup ensures entanglement propagates across the entire qubit system. In each layer \( l \) within \( n_{\text{layers}} \), the qubits are entangled following this cyclic pattern of CNOT operations:
\begin{equation}
      \text{CNOT}(q_i, q_{(i+1) \mod n_{\text{qubits}}}) \quad \text{for} \quad i = 0, 1, \ldots, n_{\text{qubits}}-1.
\end{equation}

This configuration establishes a continuous chain of entanglement, facilitating quantum information transfer across all qubits.

Furthermore, trainable quantum layers refine these representations through additional rotation operations, where each trainable weight $w_{l,i}$ is applied as a rotation:
\begin{equation}
R_Y(w_{l,i}) = \exp\left(-i w_{l,i} \frac{\sigma_y}{2}\right).
\end{equation}
This enables the model to learn optimal transformations specific to EEG encoding, allowing quantum circuits to act as feature refiners beyond conventional classical networks.

\subsection{Measurement and Classification}
After quantum transformations $U(x)$ and $V(\theta)$, qubit states are measured to extract classical information for downstream classification. Here, we choose a Pauli-Z matrix $\sigma_z$ in Eq.~(\ref{E: Pauli}) so that the final VQC output in Eq.~(\ref{E: output}) yields $\braket{\sigma_z} =(y_1, \ldots, y_n)$.

The above measurement is then converted into classical probabilities via
\begin{equation}
P(y=j | x) = \frac{\exp(y_j)}{\sum_{k} \exp(y_k)}, \quad (j=1,\ldots, n)
\end{equation}
to form a vector for classification. The results are passed into a fully connected neural network layer, where the softmax function computes the final output, producing a probability distribution over output classes for accurate EEG-based categorization.


\subsection{Computational Complexity Analysis of EEGNet and QEEGNet}

EEGNet primarily consists of convolutional layers, pooling layers, and fully connected layers. Given an input EEG signal with $C$ channels and $T$ time steps, EEGNet applies depthwise separable convolution, which has a computational complexity of
\begin{equation}
O(C \cdot T \cdot K + C \cdot M \cdot T),
\end{equation}
where $K$ represents the kernel size and $M$ is the number of output features. The first term accounts for depthwise convolution applied independently to each channel, while the second term represents the pointwise convolution operation. The overall complexity remains linear with respect to $C$ and $T$, making EEGNet computationally efficient for EEG signal processing. QEEGNet builds upon EEGNet by introducing a quantum encoding layer, which 
maps extracted EEG features onto a quantum Hilbert space using parameterized quantum circuits. The encoding step involves applying rotation-Y (RY) gates to transform classical inputs into quantum states. This quantum encoding has a computational complexity of $O(n)$, where $n$ is the number of qubits, typically equal to the number of EEG features. After encoding, QEEGNet introduces entanglement between qubits using controlled-NOT (CNOT) gates. The complexity of applying CNOT gates in a ring topology scales as $O(n)$, as each qubit interacts with a fixed number of neighbors. The measurement operation is linear with respect to the number of qubits, adding an additional $O(n)$ complexity to the model. Overall, QEEGNet introduces a modest increase in computational complexity compared to EEGNet. The total complexity of QEEGNet can be expressed as 
\begin{equation}
O(C \cdot T \cdot K + C \cdot M \cdot T + n), 
\end{equation}
where the additional $O(n)$ term accounts for quantum operations. Since $n$ typically corresponds to the number of extracted EEG features, this increase is not significant in practical scenarios. Thus, QEEGNet enhances feature representation while maintaining computational efficiency comparable to EEGNet. The integration of quantum layers provides richer feature embeddings, potentially improving EEG classification accuracy without introducing prohibitive computational overhead.

\section{Experiment Results}
\subsection{Datasets}
We tested our QEEGNet’s performance using four datasets from different sources and downstream tasks. A simple comparison of the datasets is shown in Table~\ref{tab:eegdata}.

\begin{table}[h]
    \centering
    \caption{Datasets for QEEGNet downstream tasks. MI notes Motor Imagery and ERN notes Error-Related Negativity.}
    \begin{tabular}{l l c c c}
        \hline
        \textbf{ } & \textbf{Datasets} & \textbf{Paradigms} & \textbf{Subjects} & \textbf{Classes} \\  
        \hline
        & BCIC-2A & MI & 9 & 4 \\
        & BCIC-2B & MI & 9 & 2 \\
        & KaggleERN & ERN & 26 & 2 \\
        & PhysioP300 & P300 & 9 & 2 \\
        \hline
    \end{tabular}
    \label{tab:eegdata}
\end{table}

\subsubsection{BCIC-IV-2a Dataset}
We utilize the BCIC-IV-2a dataset \cite{BCICompetition2008Graz} from the Brain-Computer Interface Competition IV, which provides time-asynchronous EEG data. As one of the most widely used public EEG datasets, it was introduced during BCI Competition IV in 2008 \cite{tangermann2012review}. The dataset comprises EEG recordings from nine subjects performing a four-class motor imagery task, which was conducted on two separate days. In each trial, participants were instructed to imagine one of four movements—right hand, left hand, feet, or tongue—for four seconds following a cue. Each session contains 288 trials, with 72 trials per movement type. The EEG data was captured using 22 electrodes positioned around the central region of the scalp at a sampling rate of 250 Hz. To preprocess the signals, we down-sampled the data to 128 Hz, applied a band-pass filter (4–38 Hz), and segmented the EEG recordings from 0.5 to 4 seconds post-cue, yielding 438 time points per trial.

\subsubsection{BCIC-IV-2b Dataset}
This dataset \cite{BCICompetition2008GrazB} consists of EEG recordings from nine right-handed subjects participating in a motor imagery experiment. Each subject completed five sessions, with two training sessions conducted without feedback and three additional sessions incorporating feedback mechanisms. EEG data was acquired from three bipolar channels (C3, Cz, C4) at a sampling rate of 250 Hz, while EOG recordings were included to assist in artifact reduction. The experimental protocol required subjects to mentally simulate left or right hand movements in response to visual cues. Each session comprised multiple runs and trials, beginning with an EOG artifact recording phase to evaluate the impact of eye movements on EEG signals. This was followed by the motor imagery task, where participants engaged in motor imagination without physical execution. This dataset serves as a valuable resource for motor imagery and brain-computer interface (BCI) research, particularly in areas related to signal processing, artifact handling, and EEG-based classification models.

\subsubsection{KaggleERN (The BCI Challenge@NER 2015) Dataset}
The Kaggle error-related negativity dataset (KaggleERN) \cite{bci_ner_2012} originates from a study involving 26 subjects who participated in a P300 Speller task using 56 EEG electrodes. This brain-computer interface (BCI) paradigm is designed to detect errors during a spelling task based on EEG signals, which were sampled at 600 Hz and referenced to the nose. The P300-Speller leverages the P300 response, an event-related potential (ERP) triggered by rare and attended stimuli, to identify the target letter from a matrix of 36 characters (letters and numbers) displayed on a computer screen. In this experiment \cite{bci_ner_2012}, participants were instructed to spell words by focusing on a target letter while ignoring irrelevant flashes. The objective was to determine whether the selected letter was incorrect by analyzing EEG signals recorded after feedback was provided. Regardless of whether the feedback was correct or incorrect, participants proceeded to the next target letter in the sequence. The study incorporated two experimental conditions: a fast mode, where each character was flashed four times, leading to a higher likelihood of errors, and a slower mode, where each character was flashed eight times, reducing error rates. Each subject completed five copy spelling sessions. The first four sessions involved spelling twelve 5-letter words, while the fifth session required spelling twenty 5-letter words. This dataset is particularly valuable for research on error detection in BCI systems, as well as studies on EEG-based classification, event-related potentials, and adaptive spelling interfaces.

\subsubsection{PhysioP300 Dataset}
The PhysioP300 dataset \cite{goldberger2000physiobank} is a publicly available EEG dataset designed for brain-computer interface (BCI) research based on P300 event-related potentials (ERPs). It includes data from nine subjects, each participating in a P300 matrix spelling paradigm. The dataset contains two target categories, typically corresponding to target and non-target stimuli, allowing for classification of brain responses to attended versus unattended visual cues. The experimental design follows the classical P300 matrix speller paradigm, where subjects focus on specific target characters within a 6×6 character matrix, while their brain responses to flashing stimuli are recorded. EEG signals were collected using a high-density EEG system, covering multiple electrodes to capture the spatial distribution of the P300 component. The stimuli were categorized into target and non-target types, with target stimuli inducing distinct P300 ERP peaks in the EEG signals. In terms of data characteristics, the dataset provides high-quality annotations, including precise event timing and target labels, facilitating temporal alignment and classification tasks. The P300 component, an exogenous evoked potential, typically appears 250–500 milliseconds after stimulus onset, with an amplitude of approximately 3–10 $\mu V$. Its spatial distribution is primarily concentrated in the parietal region, particularly around the Pz electrode. The dataset is well-suited for developing and validating P300-based BCI decoding algorithms, commonly used in classification tasks to distinguish between target and non-target stimuli. It also serves as a valuable resource for neurosignal processing algorithm validation, particularly in evaluating the performance of machine learning models for ERP classification. Additionally, cross-dataset studies frequently integrate PhysioP300 with other ERP datasets for transfer learning and meta-analysis.

\subsection{Experiment Details}
For the BCIC-IV-2a dataset, we utilized the first session of each subject as the training set, reserving one-fifth of it for validation same as \cite{chen2024qeegnet}. All the datasets we used default train-test splits. 

The data splitting strategy in the BCIC-IV-2a and BCIC-IV-2b dataset follows a cross-subject paradigm designed to evaluate model generalization across different subjects. The process begins by designating a target subject's data as the test set, which comprises all recordings from both their sessions. For the training and validation sets, the code aggregates data from all other subjects, carefully excluding the target subject to maintain proper evaluation isolation. Each contributing subject's session data undergoes a stratified split, where 90\% of the trials are allocated to the training set and 10\% to the validation set, ensuring consistent class distribution across splits. This stratified approach helps maintain balanced representation of different task conditions in both training and validation phases. The data can optionally undergo several preprocessing steps, including Ensemble Agreement processing, temporal interpolation to a specified length, and channel selection. This cross-subject validation approach is particularly valuable for assessing how well the model can adapt to new, unseen subjects, which is crucial for real-world BCI applications where model generalization across different individuals is essential. 

The KaggleERN dataset employs a structured 4-fold cross-validation approach, where each fold maintains a consistent set of test subjects while rotating different training subject combinations. Unlike the BCIC2A/2B datasets that use a single-subject validation paradigm, KaggleERN utilizes a predefined set of subjects ([1,3,4,5,8,9,10,15,19,25]) as the constant test group across all folds. The training data for each fold is drawn from different combinations of the remaining subjects, with each subject contributing data from five experimental sessions. The data processing pipeline reads EEG signals and event markers from CSV files, with separate handling for training and test datasets through dedicated functions. These functions process the raw data by applying temporal windowing from -0.7 to 2 seconds around events, selecting relevant EEG channels, and performing data normalization. This structured approach ensures consistent evaluation across folds while maintaining a clear separation between training and testing subjects, facilitating robust assessment of the model's generalization capabilities across different individuals.

The PhysioP300 dataset employs a Leave-One-Subject-Out (LOSO) cross-validation approach for evaluating model performance. The dataset comprises recordings from seven subjects (3,4,5,6,7,9,11), and the validation process iterates through each subject. In each iteration, one subject's data is held out as the validation set while the remaining subjects' data form the training set. This LOSO approach ensures a robust evaluation of the model's ability to generalize across different subjects, as it tests the model's performance on completely unseen subjects during training. The strategy is particularly valuable for assessing the practical applicability of the model in real-world scenarios where it needs to work with new users without additional training.

All the models that achieved the lowest validation loss within 100 epochs was subsequently evaluated on the second session of the same subject. 
All the models are trained with 4 qubits and 2 quantum layers expect BCIC-IV-2a used the same settings in \cite{chen2024qeegnet}. All models were implemented using PyTorch and PennyLane, with a quantum simulator serving as the backend. Training was conducted with a batch size of 32, running for 100 epochs using the AdamW optimizer with a learning rate of $10^{-3}$. The model with the highest validation accuracy was selected for final predictions on the test dataset.
The training process for QEEGNet took approximately 20 hours per subject when running on CPUs, using a Google Cloud Platform a2-ultragpu-1g machine equipped with 170 GB of RAM.

\subsection{Results}
The experimental results compare the performance of EEGNet and QEEGNet across multiple tasks and datasets, including BCIC-IV-2a, BCIC-IV-2b, KaggleERN, and PhysioP300, evaluating key metrics such as accuracy, F1-score, precision, and recall. The details are shown in Table~\ref{tab:eegbcic2a} and Table~\ref{tab:qeegother}.

For the BCIC-IV-2a dataset, QEEGNet demonstrates a slight improvement over EEGNet in both validation and test accuracy. With a validation accuracy of 42.1\% compared to EEGNet’s 39.86\%, and a test accuracy of 38.1\% versus 37.7\%, QEEGNet shows a modest but consistent enhancement in classification performance. These results suggest that QEEGNet’s quantum-inspired modifications contribute to better generalization.

On the BCIC-IV-2b dataset, EEGNet achieves a marginally higher accuracy (73.21\%) than QEEGNet (72.62\%), indicating that for this dataset, the conventional architecture of EEGNet may still hold advantages. However, QEEGNet slightly surpasses EEGNet in F1-score (70.68\% vs. 70.61\%), suggesting that it maintains a balanced trade-off between precision and recall. For precision, EEGNet outperforms QEEGNet (75.63\% vs. 73.32\%), indicating fewer false positives. However, QEEGNet achieves better recall (71.89\% vs. 69.94\%), which suggests improved sensitivity to correctly identifying relevant signals.

On the KaggleERN dataset, QEEGNet consistently outperforms EEGNet across almost all metrics. It achieves a higher accuracy of 70.65\% compared to EEGNet’s 70.19\%, showing improved classification performance. Notably, QEEGNet also surpasses EEGNet in F1-score (82.98\% vs. 81.79\%), and recall (99.33\% vs. 94.59\%). The significant boost in recall highlights QEEGNet’s ability to detect a larger proportion of relevant cases while maintaining a balanced precision.

For the PhysioP300 dataset, QEEGNet also exhibits superior performance. It achieves a test accuracy of 64.86\% compared to EEGNet’s 64.37\%. Similarly, QEEGNet outperforms EEGNet in F1-score (60.66\% vs. 59.90\%) and recall (60.03\% vs. 58.42\%), confirming its effectiveness in detecting true positive cases. While EEGNet slightly surpasses QEEGNet in precision (64.21\% vs. 62.32\%) the same as in KaggleERN dataset, the overall trend suggests that QEEGNet maintains a strong balance between precision and recall, leading to better generalization.

\begin{table}[h]
    \centering
    \caption{Comparison of EEGNet and QEEGNet on BCIC-IV-2a, the settings are the same as \cite{chen2024qeegnet}. }
    \begin{tabular}{ l c c }
        \hline
        \textbf{Model} & \textbf{Validation Accuracy} & \textbf{Test Accuracy} \\  
        \hline
        EEGNet & 39.86\%$\pm$0.138 & 37.7\%$\pm$0.126 \\
        QEEGNet & \textbf{42.1\%$\pm$0.142} & \textbf{38.1\%$\pm$0.127} \\
        \hline
    \end{tabular}
    \label{tab:eegbcic2a}
\end{table}

\begin{table}[h]
    \centering
    \caption{Comparison of EEGNet and QEEGNet on Various Datasets.}
    \label{tab:eegnet_qeegnet}
    \begin{tabular}{llccc}
        \toprule
        \textbf{Metric/Datasets} & \textbf{Model} &  \textbf{BCIC-IV-2b} & \textbf{KaggleERN} & \textbf{PhysioP300} \\
        \midrule
        \multirow{2}{*}{\textbf{Accuracy}}
        & EEGNet  & \textbf{73.21\%$\pm$0.088} & 70.19\%$\pm$0.010 & 64.37\%$\pm$0.049 \\
        & QEEGNet &  72.62\%$\pm$0.089 & \textbf{70.65\%$\pm$0.004} & \textbf{64.86\%$\pm$0.052} \\
        \midrule
        \multirow{2}{*}{\textbf{F1-score}}  
        & EEGNet  &  70.61\%$\pm$0.141 & 81.79\%$\pm$0.012 & 59.90\%$\pm$0.103 \\
        & QEEGNet &  \textbf{70.68\%$\pm$0.136 } & \textbf{82.98\%$\pm$0.003} & \textbf{60.66\%$\pm$0.082} \\
        \midrule
        \multirow{2}{*}{\textbf{Precision}}  
        & EEGNet  &  \textbf{75.63\%$\pm$0.076} & \textbf{72.12\%$\pm$0.009} & \textbf{64.21\%$\pm$0.046} \\
        & QEEGNet &  73.32\%$\pm$0.086 & 70.94\%$\pm$0.001 & 62.32\%$\pm$0.055 \\
        \midrule
        \multirow{2}{*}{\textbf{Recall}}  
        & EEGNet  &  69.94\%$\pm$0.215 & 94.59\%$\pm$0.045 & 58.42\%$\pm$0.164 \\
        & QEEGNet &  \textbf{71.89\%$\pm$0.197} & \textbf{99.33\%$\pm$0.009} & \textbf{60.03\%$\pm$0.119} \\
        \bottomrule
    \end{tabular}
    \label{tab:qeegother}
\end{table}

\begin{figure}[htbp]
    \centering
    \subfigure[Feature space visualization of EEGNet for Subject 9 in PhysioP300 dataset using t-SNE and UMAP.]{
        \includegraphics[width=1\linewidth]{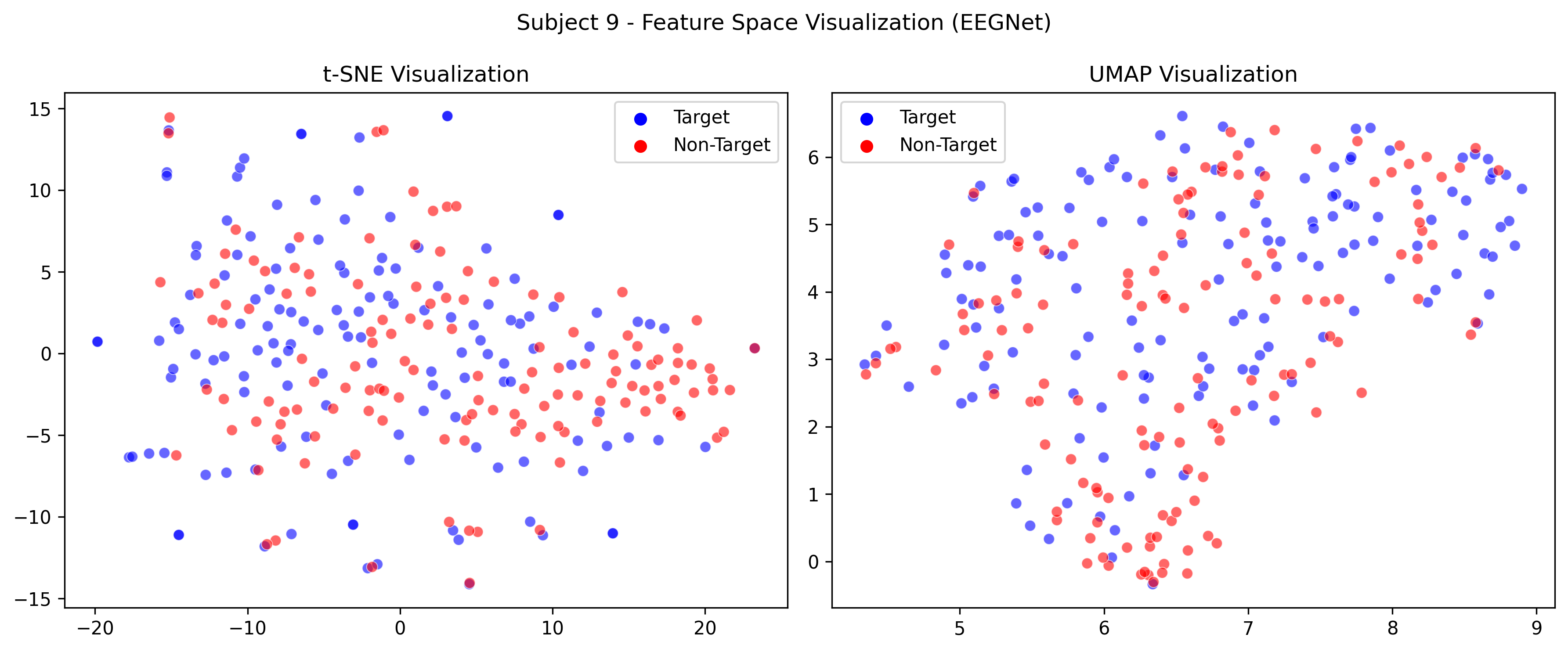}
        \label{fig:eegnet_p300}
    }
    \vspace{0.5cm} 
    \subfigure[Feature space visualization of QEEGNet for Subject 9 in PhysioP300 dataset using t-SNE and UMAP.]{
        \includegraphics[width=1\linewidth]{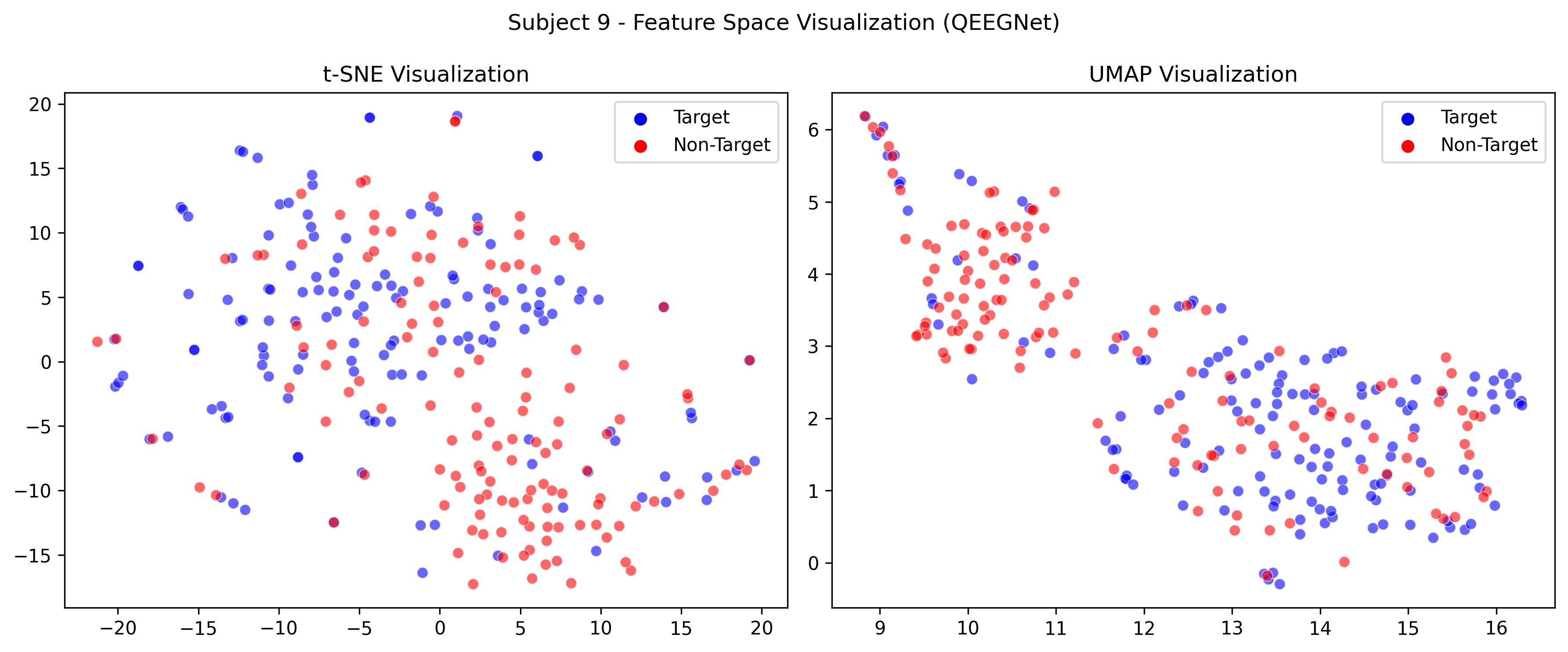}
        \label{fig:qeegnet_p300}
    }
    \caption{Comparison of feature space visualization of EEGNet and QEEGNet for Subject 9 in the PhysioP300 dataset using t-SNE and UMAP. (a) EEGNet visualization, showing the learned feature separability between target (blue) and non-target (red) samples. (b) QEEGNet visualization, demonstrating improved clustering of target and non-target samples compared to EEGNet. The clearer separation in QEEGNet’s feature space suggests its enhanced ability to distinguish P300 responses, highlighting its potential for more robust classification.}
    \label{fig:comparison}
\end{figure}

\subsubsection{Embedding Analysis}
To further evaluate the representational capability of EEGNet and QEEGNet, we visualize the learned feature embeddings using t-SNE \cite{van2008visualizing} and UMAP  \cite{mcinnes2018umap} for Subject 9 in the PhysioP300 dataset. Figure~\ref{fig:comparison} illustrates the projected feature spaces, where blue points represent target responses and red points represent non-target responses.

The t-SNE and UMAP visualizations for EEGNet (Figure 4a) indicate a moderate degree of separation between target and non-target samples, but the distributions exhibit considerable overlap. This suggests that EEGNet’s feature representations do not fully disentangle the two classes, potentially leading to misclassifications. In contrast, QEEGNet’s feature space (Figure 4b) shows a more distinct separation between target and non-target samples, particularly in the UMAP projection. The improved clustering structure in QEEGNet suggests that its learned embeddings capture more discriminative information, allowing for better differentiation of EEG responses.

The enhanced separability observed in QEEGNet's embedding space implies that it learns a more structured and class-specific representation of the EEG signals. This aligns with the improved classification performance observed in the previous results, further supporting the hypothesis that QEEGNet's quantum-inspired architecture facilitates more effective feature extraction. The clearer distinction between target and non-target samples suggests that QEEGNet has a greater potential for robust classification in EEG-based tasks.

In summary, QEEGNet demonstrates competitive and, in many cases, superior performance compared to EEGNet, particularly in recall and F1-score. While EEGNet retains advantages in precision in some datasets, QEEGNet’s improvements in accuracy and recall indicate that its quantum-inspired components contribute positively to EEG classification tasks. The results suggest that QEEGNet is particularly beneficial in scenarios where sensitivity to relevant signals is crucial, making it a promising alternative to conventional EEG decoding models.

\section{Conclusion}
This study explores the potential of integrating quantum machine learning into EEG signal processing through QEEGNet, a hybrid model that extends EEGNet with quantum layers. Our experiments across multiple datasets indicate that while QEEGNet demonstrates improvements in classification performance, particularly in recall and feature space separability, its overall gains over conventional neural networks remain moderate for the tested datasets. This suggests that the quantum-enhanced model does not yet universally outperform classical deep learning approaches in EEG decoding.

Despite these limitations, one key advantage of QEEGNet lies in its ability to achieve performance gains with only a linear increase in computational complexity. Unlike many quantum models that introduce significant computational overhead, QEEGNet maintains efficiency, making it a viable approach for practical applications. This finding highlights that quantum-inspired modifications can enhance EEG feature extraction at a relatively low computational cost, which is a promising direction for future research.

As an early-stage attempt at applying quantum computing to EEG classification, this study provides valuable insights into the feasibility and challenges of hybrid quantum-classical architectures. While further optimization is needed to fully leverage quantum advantages, our results suggest that QEEGNet represents a meaningful step toward incorporating quantum machine learning into neurotechnology and brain-computer interfaces. Future work should explore refined quantum circuit designs, better integration with classical networks, and larger-scale evaluations to further validate the potential of quantum-enhanced EEG analysis.

\bibliography{sn-bibliography,bib/vqc,bib/qml_examples}

\end{document}